\documentclass[9pt,twocolumn,twoside]{pnas-new}

\templatetype{pnasresearcharticle} 

\title{A perfect X-ray beam splitter and its applications to time-domain interferometry and quantum optics exploiting free-electron lasers}

\author[a]{\href{https://orcid.org/0000-0002-7797-6736}{Sven Reiche}}
\author[a]{\href{https://orcid.org/0000-0002-0786-3402}{Gregor Knopp}} 
\author[a]{\href{https://orcid.org/0000-0003-4228-7874}{Bill Pedrini}}
\author[a]{\href{https://orcid.org/0000-0002-7065-7417}{Eduard Prat}}
\author[a,b,c,*]{\href{https://orcid.org/0000-0001-9980-2270}{Gabriel Aeppli}}
\author[a,*]{\href{https://orcid.org/0000-0002-5717-2626}{Simon Gerber}}

\affil[a]{Paul Scherrer Institut, CH-5232 Villigen PSI, Switzerland}
\affil[b]{Department of Physics, ETH Zurich, CH-8093 Zurich, Switzerland}
\affil[c]{Institut de Physique, EPF Lausanne, CH-1015 Lausanne, Switzerland}

\leadauthor{Reiche} 

\significancestatement{Brilliant, ultrashort and coherent X-ray FELs pulses allow unique investigations of dynamics at the inherent time and length scale of atoms. However, missing are sequences of phase-locked X-ray pulses, as required for time-domain correlation spectroscopies and coherent quantum control. Based on selective electron bunch degradation in the accelerator, combined with two-stage self-seeded photon emission, we propose an FEL mode, generating sub-fs, phase-locked X-ray pulses pairs with up to 100 fs delay. Splitting the electron bunch in the accelerator, instead of photon pulses in the beamline, avoids relative phase jitter. Time-domain interferometry, such as the X-ray analog of the ubiquitous FTIR spectrometer, is enabled and, more generally, all of non-linear and quantum optics for which coherent copies of beams are required.}

\authorcontributions{S.G. and G.A. led the project. S.R. and E.P. simulated the beam dynamics. All authors contributed to the manuscript.}
\authordeclaration{Prof.~Hastings and G.A., who have never had a joint research project, participated in a workshop which resulted in a report (available as arXiv:1903.09317) co-authored by all participants, and were co-authors of an invited perspective paper on accelerator-based light sources which reported no original research (available as doi.org/10.1142/S1793626819300044). Prof.~Hastings has served on advisory panels at PSI and G.A. has served on advisory panels at SLAC where Prof.~Hastings is employed. S.R. was a postdoctoral fellow and assistant/associate researcher of Prof.~Pellegrini at UCLA from 2000 to 2008. S.G. was a postdoctoral fellow at Stanford University and SLAC from 2013 to 2015 which has led to one shared publication with Prof.~Hastings (available as doi.org/10.1073/pnas.1612849113).}
\equalauthors{\textsuperscript{*}Correspondence should be addressed to Gabriel Aeppli (+41 56 310 4232, gabriel.aeppli@psi.ch) and Simon Gerber (simon.gerber@psi.ch, +41 56 310 3965), Paul Scherrer Institut, Forschungsstr.~111,
5232 Villigen PSI,
Switzerland.}


\begin{abstract}
X-ray free-electron lasers (FEL) deliver ultrabright X-ray pulses, but not the sequences of phase-coherent pulses required for time-domain interferometry and control of quantum states. For conventional split-and-delay schemes to produce such sequences the challenge stems from extreme stability requirements when splitting Ångstrom wavelength beams where tiniest path length differences introduce phase jitter. We describe an FEL mode based on selective electron bunch degradation and transverse beam shaping in the accelerator, combined with a self-seeded photon emission scheme. Instead of splitting the photon pulses after their generation by the FEL, we split the electron bunch in the accelerator, prior to photon generation, to obtain phase-locked X-ray pulses with sub-femtosecond duration. Time-domain interferometry becomes possible, enabling the concomitant program of classical and quantum optics experiments with X-rays. The scheme leads to new scientific benefits of cutting-edge FELs with attosecond and/or high-repetition rate capabilities, ranging from the X-ray analog of Fourier transform infrared spectroscopy to damage-free measurements.
\end{abstract}

\dates{This manuscript was compiled on \today}
\doi{\url{doi/XXXXXXXXXX}}

\begin{document}

\maketitle
\thispagestyle{firststyle}
\ifthenelse{\boolean{shortarticle}}{\ifthenelse{\boolean{singlecolumn}}{\abscontentformatted}{\abscontent}}{}

\dropcap{F}or a decade, ultrashort, intense and coherent X-ray pulses for experiments in all fields of natural science have been delivered by X-ray free-electron lasers (FEL) \cite{Seddon17}. Missing are phase-locked pulses, i.e. pulse pairs with a fixed phase relation, which would allow full exploitation of the coherence properties and, thereby, extend coherent control schemes to shorter than optical and ultraviolet wavelengths \cite{Ramsey50,Mukamel95,Cundiff03,Greenland10,Kampfrath13,Chatterjee16,Wituschek20}. Foremost, multiple coherent pulses are required for many nonlinear X-ray spectroscopies. In addition, they would also enable linear schemes, such as time-domain X-ray interferometry (XRI) \cite{Chatterjee16}. Conceptually, these techniques are based on quantum interference of two photon fields with a variable time delay {$\Delta t$} and relative phase shift $\Delta\phi$. Spectral resolution is achieved by Fourier analysis and only limited by the maximum time delay at which interference is still measurable. 

Generation of phase-locked high-energy photon pulses has been proposed theoretically \cite{Thompson08} and demonstrated in the extreme ultraviolet \cite{Gauthier16,Wituschek20}. Here we introduce an FEL operation mode, where high spectral resolution is obtained from phase-locked ultrafast X-ray (PHLUX) pulses which can be delayed by up to 100~fs. The scheme is applicable in the soft and hard X-ray regime, does not require elaborate spectrometer design and allows for tuning of acquisition efficiency versus energy resolution via the variable time delay and pulse duration. Compared to other schemes (see SI Appendix), advantages are the availability of larger time delays, as well as tunability of the relative phase shift $\Delta\phi$ and amplitude of the pulses. This opens the door to coherent control and readout of prepared states, as well as the possibility of damage-free resonant X-ray scattering using $\pi$-shifted pulse pairs.

\section*{Generation of phase-locked X-ray pulses}

\begin{figure}
	\includegraphics[width=\linewidth]{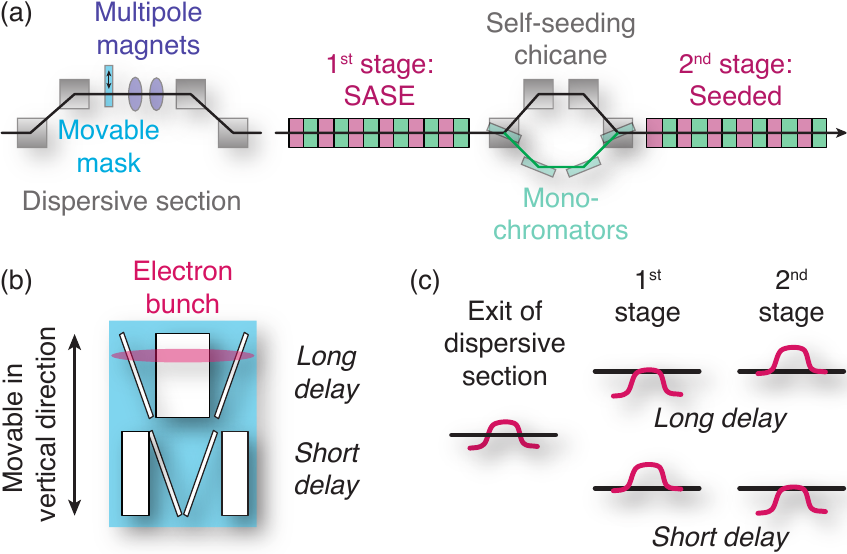}{}
	\caption{\label{layout}
		(a) X-ray FEL beamline layout with a movable micro-fabricated mask (blue) and higher-order multipole magnets (purple) in a dispersive section, as well as the two undulator sections (pink-green) that are separated by a self-seeding chicane (grey-green). (b) The movable mask features a set of slots: one set preserves the electron bunch (pink) for SASE generation, whereas two narrow slits define the unspoiled parts of the electron bunch from which the coherent signal originates. (c) The electron bunch is shaped with nonlinear transverse tilts using multipole magnets and re-aligned in the undulator section: for short (long) time delays it is aligned on axis with the tails (central part) in the first and the central part (tails) in the second stage.}
\end{figure}

Our approach to produce phase-locked pulses in the soft and hard X-ray regime is to overlay a fully coherent signal on an electron bunch that only lases in two well-defined longitudinal slices. It is achieved with a new method (see Fig.~\ref{layout}) that combines a ‘slotted’ foil \cite{Emma04,Ding15} (or,  equivalently, energy modulation or laser slicing which may be more appropriate for high repetition rate FELs), transverse tilting of the electron beam \cite{Guetg15,Guetg18,Dijkstal20}, and self-seeding \cite{Feldhaus97,Saldin01,Amann12,Ratner15,Emma17}. The latter is a two-stage process to generate coherent X-ray pulses: The first FEL stage consists of a standard self-amplified spontaneous emission (SASE) section, whose output is spectrally filtered by a monochromator. This signal is then used as a ‘seed’ for the second stage and overlapped with the ‘sliced’ parts of the electron bunch. In our method, the electron beam contains two parts defined by the slotted foil: one part is unspoiled, while the other part is spoiled except for the two regions, as defined by the slits on a micro-fabricated mask [see Fig.~\ref{layout}(b)]. We separate the two regions transversely and correct the global beam trajectory such that one region is aligned to the SASE and the other region to the self-seeding stage. As a result, the first part drives the SASE section in saturation, while the two unspoiled slices in the second region amplify the seed signal to produce a phase-locked pulse pair.

The required spatial separation of the two regions is achieved by imposing a transverse tilt on the electron bunch in the dispersive section (see Fig.~\ref{layout}(c) and the start-to-end simulations in the SI Appendix). Thereby, in both FEL stages either the central part or the tails of the bunch is/are aligned to the undulator axis, whereas the respective other part undergoes betatron oscillations and does not contribute to the lasing process. The beam tilt is imposed where also the micro-fabricated mask is placed. Multipole magnets in the dispersive section can be used to alter the longitudinal and transverse position of the electron bunch which is then preserved downstream. Namely, quadrupole magnets result in linear, sextupoles in quadratic, octupoles in cubic displacements etc. A combination of a sextupole and a decapole magnet yields the desired step function profile indicated in Fig.~\ref{layout}(c). To switch between short and long time delays amongst the phase-locked pulses, the mask is moved vertically and the alignment of the electron tilt is flipped.

\begin{figure}[t]
	\includegraphics[width=\linewidth]{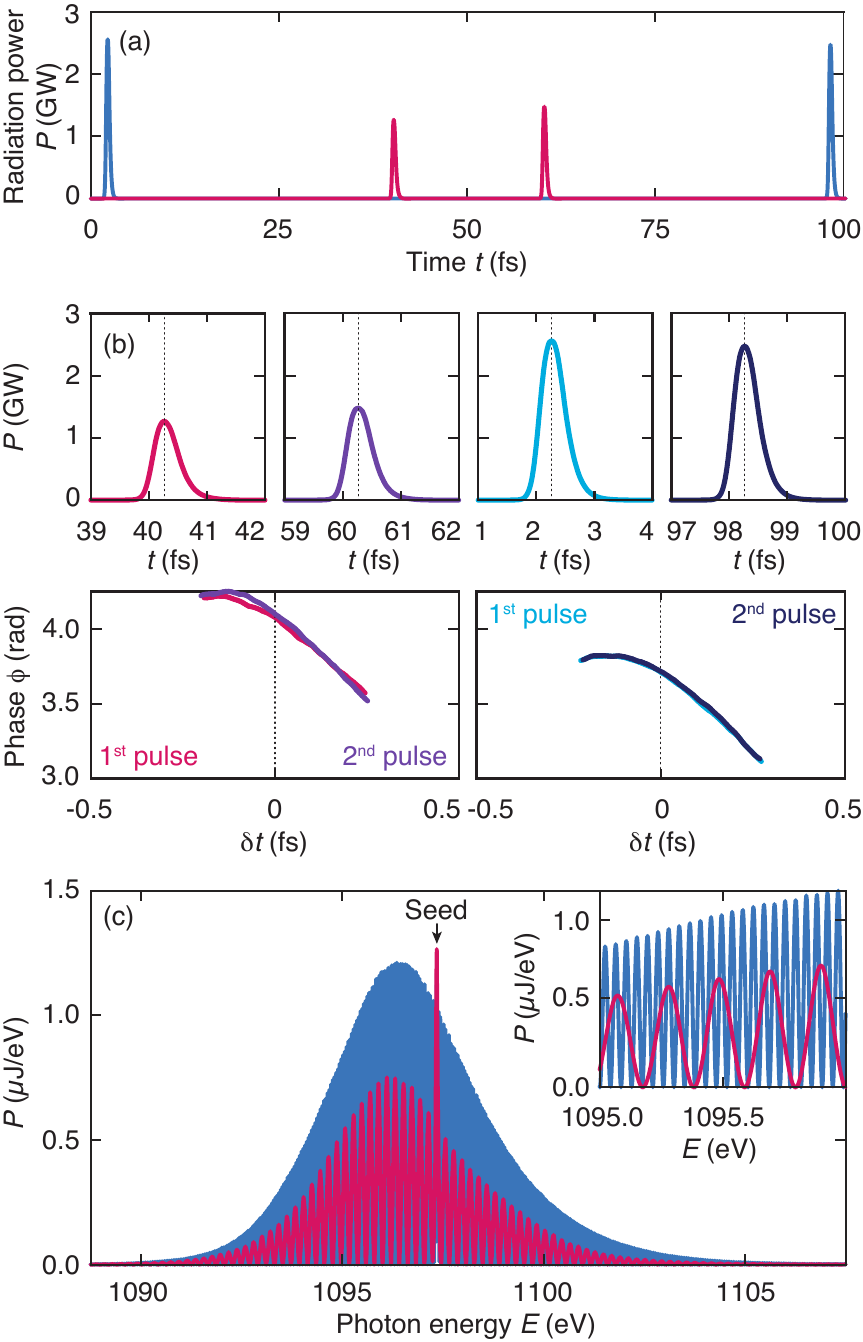}
	\caption{\label{pulses}
		 (a) Radiation power profile of pulse pairs with a time delay of $\Delta t=20$~fs (pink) and 96~fs (blue) at a seed energy of $E\sim1097$~eV. (b) Close-up view of the radiation power $P$ (upper panels) and phase $\phi$ (lower panels). Individual pulses have a duration of 0.5 fs FWHM. The phase is evaluated with respect to the central frequency of the seed. (c) Corresponding radiation power spectrum with tunable interference fringes which can be exploited for spectroscopy.}
\end{figure}

Figure~\ref{pulses} shows the performance of the PHLUX mode, in terms of radiation profile and spectrum, as well as phase stability given the baseline capabilities of the SwissFEL Athos soft X-ray branch \cite{Abela19}, assuming a seed energy of $E\sim1097$~eV and a self-seeding chicane with a resolving power of 50’000, foreseen as a future upgrade of the beamline. Here the PHLUX mode is benchmarked in the soft X-ray regime, but the concept is equally applicable to hard X-rays. To create a phase-stable wave train the bandwidth of the monochromators must be significantly smaller than the spectral width (inverse length) of individual SASE spikes (see discussion on self-seeding in the SI Appendix). The modal structure in the spectrum depends on the temporal separation of the slices and, at the maximum $\Delta t_{\rm max}=96~$fs considered here, corresponds to an energy resolution of $1/ \Delta t_{\rm max}\sim45$~meV assuming Fourier-limited pulses; a smaller self-seeding resolving power such as $\sim5$’000 implemented at LCLS \cite{Ratner15} would still give a useful $\Delta t_{\rm max}\lesssim20~$fs. A peak radiation power of $\sim2$~GW results in a peak photon field strength of $\sim1$~MV/cm at the source point. To reach non-linear driving regimes this field strength can be further increased, e.g. the baseline focusing capabilities of SwissFEL Athos allow for an increase by a factor of $\sim50$ at the sample position.

The total width of the spectrum, i.e. the number of modes, depends on the slit width on the mask---wider slits result in narrower spectra. PHLUX delivers a minimal pulse separation of $\Delta t\sim 2$~fs, which is limited by the transverse size of the electron beam. In turn, it also determines the slit width of the mask, i.e. the minimal pulse duration. We note that 2~fs is the full-width-at-half-maximum (FWHM) of the electron slice, whereas lasing occurs mainly from the central portion yielding a shorter photon pulse length of 0.5~fs FWHM. Our scheme is rather insensitive to the accuracy of the micro-fabrication process: Realistic tolerances, of order $\mu$m, result in pulse length changes of less than 100~as and have negligible effect on the phase and amplitude relation of the pulses.

\section*{Stability of phase-locked X-ray pulses}

Because the two pulses follow a common path, the PHLUX scheme is fundamentally different and offers much higher phase stability than conventional X-ray split-and-delay approaches \cite{Osaka17,Zhu17,Lu18}, where slight vibrations in the delaying monochromators and mirrors translate into phase jitter between the two pulses: For a central photon energy of $E\sim1097$~eV, $\Delta\phi=1$~ rad corresponds to a path length difference of only $\sim1.8$~Å. This translates into stability unachievable for standard setups, typically consisting of eight optical elements which need to be set and held in place with respect to each other. In contrast, Fig.~\ref{pulses}(b) shows that the peak-to-peak phase stability of PHLUX is always much better than $\Delta\phi\sim60$~mrad. The absolute phase $\phi$ is not controllable, but, importantly, jitter in $\Delta t$ and $\phi$, determined by the mask stability and quality of the self-seeding monochromator, does not affect $\Delta\phi$ (see discussion on sources of beam jitter in the Materials and Methods section). That is, the FEL itself is used as a ‘perfect’ beam splitter which protects the phase difference $\Delta\phi$ from noticeable jitter. The clean two-slit interference fringes in the radiation power spectrum [see Fig.~\ref{pulses}(c)], corresponding to the delay between pulses, attest to the quality of the pulse replication.

\begin{figure}
	\includegraphics[width=\linewidth]{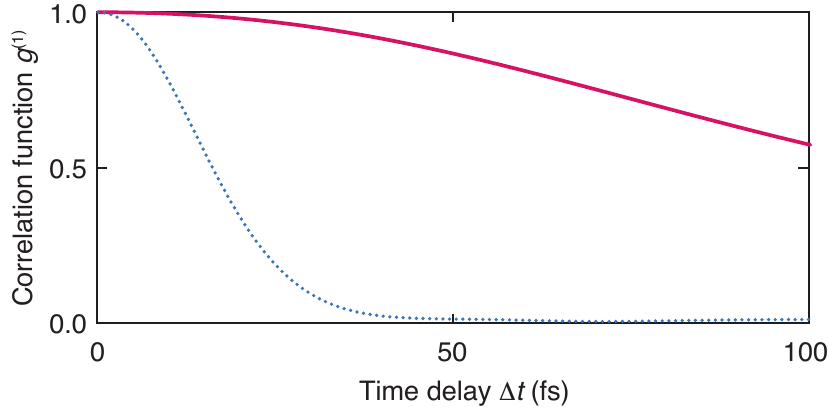}
	\caption{\label{correlation}
		Relative phase stability of the two X-ray pulses as a function of the time delay $\Delta t = t_2-t_1$, benchmarked by the first-order correlation function $g^{(1)}(t_1,t_2)$. Pink and blue lines show the results assuming a self-seeding chicane with a resolving power of 50’000 and 10’000, respectively.}
\end{figure}

The phase stability of pulse pairs can be expressed by the first-order correlation function, generally defined as
\begin{eqnarray*}
g^{(1)}(t_1,t_2)= \frac{\langle E(t_1)E^*(t_2)\rangle}{\sqrt{\langle|E(t_1)|^2\rangle\langle|E(t_2)|^2\rangle}}.
\end{eqnarray*}
$E(t)$ is  the radiation field, which is evaluated at the times $t_1$ and $t_2$=$t_1$+$\Delta t$ of the peak of the first and second pulse, respectively. $\Delta t$ is the pulse separation set by the machine parameters and the position of the slotted foil in the electron beam. Due to the phase rigidity of the individual pulses there is no need for evaluation of a further time integral. An average is taken over many shots. To calculate $g^{(1)}$ we consider that the starting signal arises due to shot-noise of the incoherent spontaneous radiation from the electron beam. Importantly, neither SASE amplification nor the self-seeding monochromators change the white-noise characteristic of this starting signal \cite{Saldin00}. Therefore, the coherence properties are defined by the resolving power of the monochromators and a Monte-Carlo evaluation permits to determine the coherence function $g^{(1)}(t_1,t_2)$. Amplification of two slices from the output field of the monochromators then inherits these coherence properties unless the pulses are driven deep into saturation, which anyhow should be avoided as it distorts the spectral quality of the FEL signal. Due to the nature of white noise, a fixed phase relation also results in a stable amplitude relation \cite{Goodman15}.

Figure~\ref{correlation} illustrates the relative phase stability as a function of $\Delta t = t_2-t_1$, reconstructed via $g^{(1)}(t_1,t_2)$. The signal degrades with increasing $\Delta t$, indicating the importance of a self-seeding chicane with highest possible resolving power. $g^{(1)}=1$ means that the relative phase difference between two pulses is stable over many shots (coherent light), while for $g^{(1)}=0$ the relative phase between pulses fluctuates randomly (chaotic light). 
A source for the reduced stability at large $\Delta t$ could, in principle, be intrinsic shot noise fluctuations, which mainly affect the pulse phase at low seed power levels ($\sim10$~kW). However, this is not a concern for the parameters used here, when the electron bunch is driven close to saturation in the first FEL stage (MW seed power, see discussion on self-seeding in the SI Appendix). On the other hand, we find rather tight but nonetheless achievable tolerances for electron beam parameters such as the current, energy, energy spread, and transverse offset. All impact the gain length and, thereby, the phase from seeding to saturation. We note that this not only forms a limitation but also provides a means to tune the phase and amplitude relation of the pulses (see Materials and Methods).

\section*{Time-domain interferometry}

\begin{figure}
	\includegraphics[width=\linewidth]{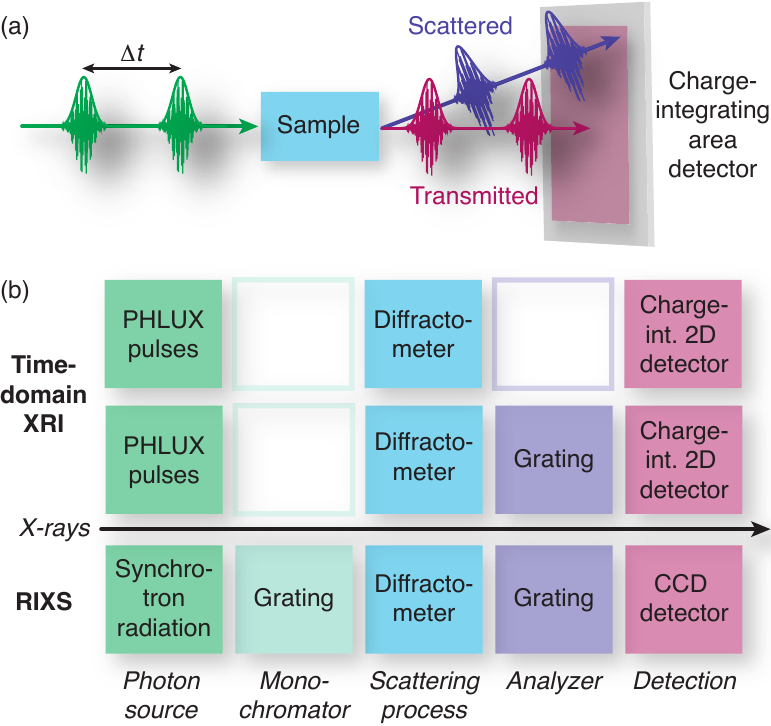}
	\caption{\label{benchmark}
		(a) Schematic of time-domain X-ray interferometry (XRI) experiments where an incident pulse pair (green) with a fixed phase relation and time delay $\Delta t$ is transmitted through (pink) or scattered by (purple) a sample. The respective signal is recorded on a charge-integrating area detector. (b) Frequency-domain RIXS requires high-resolution gratings before and after the sample for incident photon energy selection and analysis of the scattered signal. Instead, XRI does not require a monochromator after the undulators and can benefit from the multiplexing of an area detector.}
\end{figure}

We turn now to the applications of our proposed \mbox{X-ray} beam splitter. The first is to take advantage of the power spectrum with tunable modulation and phase [see Fig.~\ref{pulses}(c)] to characterize absorption lines by their Fourier transforms without moving a monochromator. Spectroscopy can be performed by varying the phase shift and/or the delay between the pulses and collecting integrated counts on a detector [see Fig.~\ref{benchmark}(a)]. This technique is in complete analogy with Fourier transform infrared (FTIR) spectroscopy where the incident beam is prepared in the spatial rather than the time domain by positioning a mirror. Such measurements---now in the X-ray regime---can underpin (optical) pump -- (X-ray absorption) probe experiments of e.g. core-hole lifetimes modified via photo-excited modulation of states at the Fermi level. Judicious choices of Fourier components, where $\Delta E$ is specified by particular values of $\Delta t$, allow for highly-efficient determinations of key parameters such as the widths of well-defined lines.

Further possibilities exploiting the modulation illustrated in Fig.~\ref{pulses}(c) and the presence of coherent processes in the sample include resonant scattering performed with and without an analyzer [see Fig.~\ref{benchmark}(b)]. The former enables resonant inelastic X-ray scattering (RIXS) where scans are performed by varying the incident beam modulation for fixed outgoing beam energies rather than by moving a monochromator, and is complementary to RIXS instrument concepts where the incident beam energy is encoded spatially on samples \cite{Strocov10,Zhou20}. Benchmarking of frequency-domain RIXS against time-domain XRI shows that the latter can be a game changer at high-repetition rate X-ray FELs (see SI Appendix). 

Given that we are working with coherent beams on timescales comparable to core-hole lifetimes, it is interesting and important to consider that there is a similar timescale $\tau$ for the liberation of electrons from the cores of the atoms used for photodetection. This means, in a semiclassical description, that the detector measures not simply the integral over the square of the impinging time-dependent optical field $A(t)$, but rather the integral of the square of the convolution of $A(t)$ with a probability amplitude for the release of the electrons from the cores, which we take to decay exponentially as $e^{-t/\tau}$. The detector signal is then given by
\begin{eqnarray*}
    I=\tau^{-2}\int \left| \int A(t')e^{(t-t')/\tau}\theta(t'-t)dt'\right|^2 dt,
\end{eqnarray*}
where $\theta(t)$ is a Heaviside step function. 

Our ‘beam splitter’ produces optical fields which can be regarded as sums of delta functions $\delta(t)$ in time, which can undergo stretching, or be scattered from or transmitted through a sample. Whether or not such stretching occurs, the basic physics of what is measured by the detector is still captured by assuming \mbox{$A(t)=A_1 \delta(t) + A_2 \delta(t-\Delta t)$} from which one obtains 
\begin{eqnarray*}
    I\propto|A_1|^2+|A_2|^2+ 2\operatorname{Re}(A_1A_2^*)e^{-\Delta t/\tau}.
\end{eqnarray*}

The phase coherence of the pulse pairs guarantees that over timescales set by decoherence in the detector, i.e. $\tau\gtrsim\Delta t$, there will be visible interference terms which measure directly the correlation between scattering amplitudes at different times for the sample. Thus, if we consider that the momentum transfer $q$ is defined by the position of the relevant pixel on the detector [see Fig.~\ref{benchmark}(a)], we are seeing the intermediate scattering functions $F(q,t)$ of the sample, which is proportional to the Fourier transform in space of the time-dependent two-particle correlation functions $G(r,t)$ for generalized charge and/or magnetization densities.

On the other hand, for $\tau\ll \Delta t$, we will simply see the superposition of the ordinary scattering patterns, i.e. $I\propto|A_1|^2+|A_2|^2$, and we can, by averaging over many pulse pairs, only measure the four-particle correlation function associated with speckle \cite{Grubel07}. The ergodic theorem tells us that speckle should rigorously vanish in the infinite volume limit for systems at equilibrium~\cite{Goodman15}. However, the requirements for ergodicity are hardly met for many samples of contemporary interest, and our scheme will allow speckle correlations to be measured at unprecedentedly short times. We note though that minimal speckle represents an advantage for isolating the interference terms probing the two-particle correlation functions, and for following these to $\Delta t > \tau$. 

We turn now to what may well be the most far-reaching implication of phase-locked pairs of sub-fs X-ray pulses: the possibility of implementing the full program of quantum optics with X-rays. In particular, our scheme permits tuning of the phase difference and amplitudes of the two pulses (see Materials and Methods), enabling coherent control and readout of prepared states, e.g. in photon-echo type experiments, which at FELs have been demonstrated in the far infrared \cite{Greenland10,Litvinenko15,Chick17} and extreme ultraviolet regime \cite{Wituschek20}. The latter builds on a combination of phase-modulated seed pulses and high-gain harmonic generation with which Ramsey fringes have been detected up to $E=47.5$~eV. However, this approach cannot delay pulses less than $\Delta t\sim150$~fs, which is long compared to the decoherence times $\tau$ for excitations of atomic cores (that is also relevant for the detector as described above). In contrast, with the PHLUX scheme pulses are `split' in the electron accelerator which permits application at higher (soft and hard X-ray) energies and provides access to a larger momentum range. The addition of multiple evenly-spaced slits on the micro-fabricated mask is an extension of the design shown in Fig.~\ref{layout}(b) and yields trains of phase-stable sub-fs pulses---an X-ray frequency comb.

Importantly, using our mode $\Delta t$ can also be reduced to a few femtoseconds. In combination with the tunability of $\Delta\phi$ this allows for two few-fs delayed X-ray pulses with a phase difference of $\Delta\phi=\pi$. Beyond a critical threshold intensity, self-induced transparency occurs when short coherent light pulses interact with a dense medium, resulting in anomalously low absorption~\cite{McCall67,McCall69,Stohr15}. Namely, this holds when, within an excitation cycle, the same amount of energy is coherently absorbed by a resonant two-level system, as is coherently emitted thereafter. For the single pulse experiments performed to date scattering is eliminated along with absorption \cite{Wu16,Chen18}. PHLUX enables generalizations of such experiments to pairs of $\pi$-shifted pulses, where the first pulse resonantly excites and, shortly thereafter, the second pulse resonantly de-excites the sample. Such a sequence could take place on timescales faster than radiation damage occurs and, nonetheless, would allow certain scattered signals to emerge. This would represent another potential route, in addition to ghost imaging \cite{Li17}, to damage-free X-ray scattering, with profound implications for all fields of \mbox{X-ray} science, particularly also for first-principle structure determination of solids and biological samples.

\subsection*{Data Archival}

All study data are included in the article and/or SI Appendix.

\matmethods{
Simulations of the phase-locked ultrafast X-ray (PHLUX) mode were carried out with the 3D time-dependent free-electron laser (FEL) code \textit{Genesis~1.3} \cite{Reiche99}, currently released in its forth version and available at available at \href{http://genesis.web.psi.ch}{http://genesis.web.psi.ch}, using the parameters of the SwissFEL Athos soft X-ray beamline \cite{Abela19} at a seed photon energy of \mbox{$E=1097$~eV}. The input was prepared to model the emittance degradation from the mask and the seed signal. This also included variation of beam parameters for the slotted foil location to study the tolerances in the FEL performance. In addition, the particle tracker \textit{Elegant} \cite{Borland00} was used to simulate the effect of the slotted foil on the electron bunch and the tilt with the higher-order multipole magnets in the dispersive section of the beamline. After the tracking with \textit{Elegant} the sliced beam parameters were analyzed to generate the corresponding input for the \textit{Genesis} simulation.  

\subsection*{Validation of beam dynamics simulations}

\textit{Genesis 1.3} \cite{Reiche99} is one of the most commonly used codes for designing FEL facilities or to verify and interpret beam dynamics data. Since reproducing the results from the first hard X-ray lasing of LCLS \cite{Emma10} the code has been expanded to also include more advanced modelling, such as the recent echo-enabled harmonic generation experiment at FERMI \cite{Ribic19}, which is extremely challenging due to the high harmonic conversion from 260 down to 4~nm. The simulation also agrees well with the performance of the SwissFEL Aramis hard X-ray beamline \cite{Prat20}, all of which ensure our modelling capability of advanced operation modes such as PHLUX.

\subsection*{Sources of beam jitter}

We elaborate on the stability of the relative phase $\Delta\phi$ between the two pulses with respect to jitter in the electron beam properties. Unlike conventional split-and-delay methods, jitter in the time separation between the two pulses affects the PHLUX operation mode less, since $\Delta\phi$ is given by the monochromator settings (see \textit{Tuning of the phase difference and radiation power}), as well as the electron bunch phase space portrait going into the undulators after the initial seeding stage. Also, FEL amplification of the two ‘sliced’ pulses enhances the amplitude of the radiation by several orders of amplitude, but changes the radiation phase of the seed only very little. Therefore, interference of the two pulses remains constructive at the central wavelength of the seed signal. Only the modulation within the envelope of the power spectrum [see inset of Fig.~{\color{blue}2}(c)] varies with jitter but the central line remains unchanged. Moreover, there is also only a weak phase variation if the growth rate of the FEL amplification process between the two slices changes from shot to shot.
Since global jitter, such as that of the electron beam mean energy, affects both slices equally, the phase relation between the two FEL pulses remains unchanged and, therefore, the stability of the central frequency in the interference spectrum is preserved. Only relative beam parameter jitter has an effect on $\Delta\phi$. Here the strongest contribution arises from the shot-to-shot fluctuation in the electron beam parameters due to the micro bunch instability. In seeded FELs this becomes apparent in the so-called pedestal of the spectra \cite{Ratner15,Amann12,Nam21}. For our application this results in a loss of coherence between the two pulses (damping of first-order correlation function $g^{(1)}(t_1,t_2)$ shown in Fig.~3). The micro bunching instability is difficult to simulate, but its impact on the performance of PHLUX scheme can be estimated from varying the beam parameters. Figure~S1 in the SI Appendix shows this dependence of the radiation power and phase relation on the most important sources of jitter. At the SwissFEL Athos beamline we expect that the relative peak current and energy spread vary by less than 1\% and, thus, do not significantly contribute to jitter between the two pulses. Concerning the relative mean energy and trajectory jitter, the standard requirements to operate self-seeding, e.g. $<10^{-4}$ mean energy and $<10~\mu$m orbit stability, already imply acceptable phase and power jitter between the two pulses. These estimates are based on the assumption of an optimally configured laser heater \cite{Huang10} which minimizes the impact of the microbunching instability.

\subsection*{Tuning of the phase difference and radiation power}

The relative phase difference $\Delta\phi$ between the two \mbox{X-ray} pulses can be controlled by a slight detuning of the self-seeding monochromator: If the beam energy, undulator field and central wavelength of the self-seeding monochromator are identical, then the phase along the seed pulse is constant. Detuning of the later induces a linear spatial change of the seed pulse phase. Consequently, also the sliced portions of the pulse then feature a phase difference given by 
\begin{eqnarray*}
    \Delta\phi = (k_{\rm MC}-k_{\rm R}) c \Delta t,
\end{eqnarray*}
where $c$ is the speed of light. $k_{\rm MC}$ and $k_{\rm R}$ are the kinetic energies of the self-seeding monochromator and radiator, respectively. For example, a seed photon energy of \mbox{$E=1097$~eV} and time delay of $\Delta t = 7$~fs requires a relative detuning of $2.7\cdot10^{-4}$ for a pulse-to-pulse phase difference of $\Delta\phi=\pi$. Such manipulation is difficult for pulse pairs that overlap in time, but else this variant of the PHLUX scheme does not represent a major restriction in terms of operation.

The total radiation power is tunable by use of a so-called laser heater \cite{Huang10} or by removing/adding undulator modules that contribute to the lasing process. The relative amplitude or rather radiation power $\Delta P$ of the two pulses can then be varied by, for example, adding a transverse or energy chirp to the beam. That way, the radiation power of the two phase-locked pulses can be individually adjusted, e.g. for Ramsey ($\pi/2$-$\pi/2$) or Hahn-echo ($\pi/2$-$\pi$) type experiments.

}

\showmatmethods{} 

\acknow{The authors acknowledge fruitful discussions with A. Cavalieri, R. Follath, G. Matmon, L. Patthey, B.~Schmitt, T.~Schmitt and V. N. Strocov. In the course of the review process we became aware of the recent demonstration of phase-stable hard X-ray pulse pairs based on superfluorescence and seeded stimulated emission \cite{Zhang21} [available as arXiv:2110.08262 (2021)], which underlines the importance of the more general working principle and enabled experiments reported here [available as arXiv:2010.00230 (2020)]. This project received funding from the European Research Council under the European Union’s Horizon 2020 research and innovation program HERO (Grant agreement No.~810451). G.K. was supported by the Swiss National Science Foundation (Contract No.~165550).}

\showacknow{} 

\bibliography{PHLUX}

\begin{thebibliography}{10}

\bibitem{Seddon17}
EA Seddon, et~al., Short-wavelength free-electron laser sources and science: A
  review.
\newblock {\em\protect\JournalTitle{Rep. Prog. Phys.}} \textbf{80}, 115901
  (2017).

\bibitem{Ramsey50}
NF Ramsey, A molecular beam resonance method with separated oscillating fields.
\newblock {\em\protect\JournalTitle{Phys. Rev.}} \textbf{78}, 695 (1950).

\bibitem{Mukamel95}
S Mukamel, {\em Principles of Nonlinear Optical Spectroscopy}.
\newblock (Cambridge University Press), (1995).

\bibitem{Cundiff03}
ST Cundiff, J Ye, Femtosecond optical frequency combs.
\newblock {\em\protect\JournalTitle{Rev. Mod. Phys.}} \textbf{75}, 325 (2003).

\bibitem{Greenland10}
PT Greenland, et~al., Coherent control of {R}ydberg states in silicon.
\newblock {\em\protect\JournalTitle{Nature}} \textbf{465}, 1057 (2010).

\bibitem{Kampfrath13}
T Kampfrath, K Tanaka, KA Nelson, Resonant and nonresonant control over matter
  and light by intense terahertz transients.
\newblock {\em\protect\JournalTitle{Nat. Photon.}} \textbf{9}, 680 (2013).

\bibitem{Chatterjee16}
S Chatterjee, T Nakajima, Ramsey interferometry for resonant {A}uger decay
  through core-excited states.
\newblock {\em\protect\JournalTitle{Phys. Rev. A}} \textbf{94}, 023417 (2016).

\bibitem{Wituschek20}
A Wituschek, et~al., Tracking attosecond electronic coherences using
  phase-manipulated extreme ultraviolet pulses.
\newblock {\em\protect\JournalTitle{Nat. Commun.}} \textbf{11}, 883 (2020).

\bibitem{Thompson08}
NR Thompson, BWJ McNeil, Mode locking in a free-electron laser amplifier.
\newblock {\em\protect\JournalTitle{Phys. Rev. Lett.}} \textbf{100}, 203901
  (2008).

\bibitem{Gauthier16}
D Gauthier, et~al., Generation of phase-locked pulses from a seeded
  free-electron laser.
\newblock {\em\protect\JournalTitle{Phys. Rev. Lett.}} \textbf{116}, 024801
  (2016).

\bibitem{Emma04}
P Emma, et~al., Femtosecond and subfemtosecond {X}-ray pulses from a
  self-amplified spontaneous-emission-based free-electron laser.
\newblock {\em\protect\JournalTitle{Phys. Rev. Lett.}} \textbf{92}, 074801
  (2004).

\bibitem{Ding15}
Y Ding, et~al., Generating femtosecond {X}-ray pulses using an
  emittance-spoiling foil in free-electron lasers.
\newblock {\em\protect\JournalTitle{Appl. Phys. Lett.}} \textbf{107}, 191104
  (2015).

\bibitem{Guetg15}
MW Guetg, B Beutner, E Prat, S Reiche, Optimization of free electron laser
  performance by dispersion-based beam-tilt correction.
\newblock {\em\protect\JournalTitle{Phys. Rev. ST Accel. Beams}} \textbf{18},
  030701 (2015).

\bibitem{Guetg18}
MW Guetg, AA Lutman, Y Ding, TJ Maxwell, Z Huang, Dispersion-based fresh-slice
  scheme for free-electron lasers.
\newblock {\em\protect\JournalTitle{Phys. Rev. Lett.}} \textbf{120}, 264802
  (2018).

\bibitem{Dijkstal20}
P Dijkstal, A Malyzhenkov, S Reiche, E Prat, Demonstration of two-color {X}-ray
  free-electron laser pulses with a sextupole magnet.
\newblock {\em\protect\JournalTitle{Phys. Rev. Accel. Beams}} \textbf{23},
  030703 (2020).

\bibitem{Feldhaus97}
J Feldhaus, EL Saldin, JR Schneider, EA Schneidmiller, MV Yurkov, Possible
  application of {X}-ray optical elements for reducing the spectral bandwidth
  of an {X}-ray {SASE FEL}.
\newblock {\em\protect\JournalTitle{Opt. Commun.}} \textbf{140}, 341 (1997).

\bibitem{Saldin01}
EL Saldin, EA Schneidmiller, YV Shvyd’ko, MV Yurkov, X-ray {FEL} with a me{V}
  bandwidth.
\newblock {\em\protect\JournalTitle{AIP Conf. Proc.}} \textbf{581}, 153 (2001).

\bibitem{Amann12}
J Amann, et~al., Demonstration of self-seeding in a hard-{X}-ray free-electron
  laser.
\newblock {\em\protect\JournalTitle{Nat. Photon.}} \textbf{6}, 693--698 (2012).

\bibitem{Ratner15}
D Ratner, et~al., Experimental demonstration of a soft {X}-ray self-seeded
  free-electron laser.
\newblock {\em\protect\JournalTitle{Phys. Rev. Lett.}} \textbf{114}, 054801
  (2015).

\bibitem{Emma17}
C Emma, et~al., Experimental demonstration of fresh bunch self-seeding in an
  {X}-ray free electron laser.
\newblock {\em\protect\JournalTitle{Appl. Phys. Lett.}} \textbf{100}, 154101
  (2017).

\bibitem{Abela19}
R Abela, et~al., The {S}wiss{FEL} soft {X}-ray free-electron laser beamline:
  Athos.
\newblock {\em\protect\JournalTitle{J. Synchrotron Rad.}} \textbf{26}, 1073
  (2019).

\bibitem{Osaka17}
T Osaka, et~al., Characterization of temporal coherence of hard {X}-ray
  free-electron laser pulses with single-shot interferograms.
\newblock {\em\protect\JournalTitle{IUCrJ}} \textbf{4}, 728 (2017).

\bibitem{Zhu17}
D Zhu, et~al., Development of a hard {X}-ray split-delay system at the {L}inac
  {C}oherent {L}ight {S}ource.
\newblock {\em\protect\JournalTitle{Proc. of SPIE}} \textbf{10237}, 102370R
  (2017).

\bibitem{Lu18}
W Lu, et~al., Development of a hard {X}-ray split-and-delay line and
  performance simulations for two-color pump-probe experiments at the
  {E}uropean {XFEL}.
\newblock {\em\protect\JournalTitle{Rev. Sci. Instrum.}} \textbf{89}, 063121
  (2018).

\bibitem{Saldin00}
EL Saldin, EV Schneidmiller, MV Yurkov, {\em The Physics of Free Electron
  Lasers}.
\newblock (Springer-Verlag), (2000).

\bibitem{Goodman15}
JW Goodman, {\em Statistical Optics}.
\newblock (Wiley), (2015).

\bibitem{Strocov10}
VN Strocov, Concept of a spectrometer for resonant inelastic {X}-ray scattering
  with parallel detection in incoming and outgoing photon energies.
\newblock {\em\protect\JournalTitle{J. Synchrotron Rad.}} \textbf{17}, 103
  (2010).

\bibitem{Zhou20}
KJ Zhou, S Matsuyama, VN Strocov, $hv^2$-concept breaks the photon-count limit
  of {RIXS} instrumentation.
\newblock {\em\protect\JournalTitle{J. Synchrotron Rad.}} \textbf{27}, 1235
  (2020).

\bibitem{Grubel07}
G Grübel, GB Stephenson, C Gutt, H Sinn, T Tschentscher, {XPCS} at the
  {E}uropean {X}-ray free electron laser facility.
\newblock {\em\protect\JournalTitle{Nucl. Instr. and Meth. in Phys. Res. B}}
  \textbf{262}, 357 (2007).

\bibitem{Litvinenko15}
KL Litvinenko, et~al., Coherent creation and destruction of orbital wavepackets
  in {S}i:{P} with electrical and optical read-out.
\newblock {\em\protect\JournalTitle{Nat. Commun.}} \textbf{6}, 6549 (2015).

\bibitem{Chick17}
S Chick, et~al., Coherent superpositions of three states for phosphorous donors
  in silicon prepared using {TH}z radiation.
\newblock {\em\protect\JournalTitle{Nat. Commun.}} \textbf{8}, 16038 (2017).

\bibitem{McCall67}
SL McCall, EL Hahn, Self-induced transparency by pulsed coherent light.
\newblock {\em\protect\JournalTitle{Phys. Rev. Lett.}} \textbf{18}, 908--911
  (1967).

\bibitem{McCall69}
SL McCall, EL Hahn, Self-induced transparency.
\newblock {\em\protect\JournalTitle{Phys. Rev.}} \textbf{183}, 457--485 (1969).

\bibitem{Stohr15}
J Stöhr, A Scherz, Creation of {X}-ray transparency of matter by stimulated
  elastic forward scattering.
\newblock {\em\protect\JournalTitle{Phys. Rev. Lett.}} \textbf{115}, 019902
  (2015).

\bibitem{Wu16}
B Wu, et~al., Elimination of {X}-ray diffraction through stimulated {X}-ray
  transmission.
\newblock {\em\protect\JournalTitle{Phys. Rev. Lett.}} \textbf{117}, 027401
  (2016).

\bibitem{Chen18}
Z Chen, et~al., Ultrafast self-induced {X}-ray transparency and loss of
  magnetic diffraction.
\newblock {\em\protect\JournalTitle{Phys. Rev. Lett.}} \textbf{121}, 137403
  (2018).

\bibitem{Li17}
Z Li, N Medvedev, HN Chapman, Y Shih, Radiation damage free ghost diffraction
  with atomic resolution.
\newblock {\em\protect\JournalTitle{J. Phys. B: At. Mol. Opt. Phys}}
  \textbf{51}, 025503 (2017).

\bibitem{Reiche99}
S Reiche, {GENESIS} 1.3: A fully 3{D} time-dependent {FEL} simulation code.
\newblock {\em\protect\JournalTitle{Nucl. Instr. Meth. Phys. Res. A}}
  \textbf{429}, 243 (1999).

\bibitem{Borland00}
M Borland, Elegant: A flexible sdds-compliant code for accelerator simulation,
  (Advanced Photon Source), Technical Report LS-287 (2000).

\bibitem{Emma10}
P Emma, et~al., First lasing and operation of an ångstrom-wavelength
  free-electron laser.
\newblock {\em\protect\JournalTitle{Nat. Photon.}} \textbf{4}, 641 (2010).

\bibitem{Ribic19}
PR Ribič, et~al., Coherent soft {X}-ray pulses from an echo-enabled harmonic
  generation free-electron laser.
\newblock {\em\protect\JournalTitle{Nat. Photon.}} \textbf{13}, 555 (2019).

\bibitem{Prat20}
E Prat, et~al., A compact and cost-effective hard {X}-ray free-electron laser
  driven by a high-brightness and low-energy electron beam.
\newblock {\em\protect\JournalTitle{Nat. Photon.}} \textbf{14}, 748 (2020).

\bibitem{Nam21}
I Nam, et~al., High-brightness self-seeded {X}-ray free-electron laser covering
  the 3.5 ke{V} to 14.6 ke{V} range.
\newblock {\em\protect\JournalTitle{Nat. Photon.}} \textbf{15}, 435 (2021).

\bibitem{Huang10}
Z Huang, et~al., Measurements of the {L}inac {C}oherent {L}ight {S}ource laser
  heater and its impact on the {X}-ray free-electron laser performance.
\newblock {\em\protect\JournalTitle{Phys. Rev. ST Accel. Beams}} \textbf{13},
  020703 (2010).

\bibitem{Zhang21}
Y Zhang, et~al., Generation of intense phase-stable femtosecond hard {X}-ray
  pulse pairs.
\newblock {\em\protect\JournalTitle{arXiv:2110.08262}} (2021).

\end{thebibliography}


\begin{thebibliography}{10}

\bibitem{Abela19}
R Abela, et~al., The {S}wiss{FEL} soft {X}-ray free-electron laser beamline:
  Athos.
\newblock {\em\protect\JournalTitle{J. Synchrotron Rad.}} \textbf{26}, 1073
  (2019).

\bibitem{Borland00}
M Borland, Elegant: A flexible sdds-compliant code for accelerator simulation,
  (Advanced Photon Source), Technical Report LS-287 (2000).

\bibitem{Reiche99}
S Reiche, {GENESIS} 1.3: A fully 3{D} time-dependent {FEL} simulation code.
\newblock {\em\protect\JournalTitle{Nucl. Instr. Meth. Phys. Res. A}}
  \textbf{429}, 243 (1999).

\bibitem{Nam21}
I Nam, et~al., High-brightness self-seeded {X}-ray free-electron laser covering
  the 3.5 ke{V} to 14.6 ke{V} range.
\newblock {\em\protect\JournalTitle{Nat. Photon.}} \textbf{15}, 435 (2021).

\bibitem{Prat18}
E Prat, S Reiche, Compact coherence enhancement by subharmonic self-seeding in
  {X}-ray free-electron laser facilities.
\newblock {\em\protect\JournalTitle{J. Synchrotron Rad.}} \textbf{25}, 329
  (2018).

\bibitem{Ratner15}
D Ratner, et~al., Experimental demonstration of a soft {X}-ray self-seeded
  free-electron laser.
\newblock {\em\protect\JournalTitle{Phys. Rev. Lett.}} \textbf{114}, 054801
  (2015).

\bibitem{Thompson08}
NR Thompson, BWJ McNeil, Mode locking in a free-electron laser amplifier.
\newblock {\em\protect\JournalTitle{Phys. Rev. Lett.}} \textbf{100}, 203901
  (2008).

\bibitem{Beye13}
M Beye, P Wernet, C Schüßler-Langeheine, A Föhlisch, Time resolved resonant
  inelastic {X}-ray scattering: {A} supreme tool to understand dynamics in
  solids and molecules.
\newblock {\em\protect\JournalTitle{J. Electron. Spectrosc.}} \textbf{188}, 172
  (2013).

\bibitem{Ament11}
LJP Ament, M van Veenendaal, TP Devereaux, JP Hill, Resonant inelastic {X}-ray
  scattering studies of elementary excitations.
\newblock {\em\protect\JournalTitle{Rev. Mod. Phys.}} \textbf{85}, 705 (2011).

\bibitem{Strocov10}
VN Strocov, Concept of a spectrometer for resonant inelastic {X}-ray scattering
  with parallel detection in incoming and outgoing photon energies.
\newblock {\em\protect\JournalTitle{J. Synchrotron Rad.}} \textbf{17}, 103
  (2010).

\bibitem{Huang18}
DJ Huang, CT Chen, Quest for ultra-high-resolution high-efficiency resonant
  inelastic soft {X}-ray scattering.
\newblock {\em\protect\JournalTitle{Synchrotron Radiat. News}} \textbf{31}, 3
  (2018).

\bibitem{Jarrige18}
I Jarrige, V Bisogni, Y Zhu, W Leonhardt, J Dvorak, Paving the way to
  ultra-high-resolution resonant inelastic {X}-ray scattering with the {SIX}
  beamline at {NSLS-II}.
\newblock {\em\protect\JournalTitle{Synchrotron Radiat. News}} \textbf{31}, 7
  (2018).

\bibitem{Brookes18}
NB Brookes, et~al., The beamline {ID32} at the {ESRF} for soft {X}-ray high
  energy resolution resonant inelastic {X}-ray scattering and polarisation
  dependent {X}-ray absorption spectroscopy.
\newblock {\em\protect\JournalTitle{Nucl. Instr. and Meth. in Phys. Res. A}}
  \textbf{903}, 175 (2018).

\bibitem{Strocov10ii}
VN Strocov, et~al., High-resolution soft {X}-ray beamline {ADRESS} at the
  {S}wiss {L}ight {S}ource for resonant inelastic {X}-ray scattering and
  angle-resolved photoelectron spectroscopies.
\newblock {\em\protect\JournalTitle{J. Synchrotron Rad.}} \textbf{17}, 631
  (2010).

\bibitem{Dunn18}
M Dunn, RW Schönlein, Future directions of high repetition rate {X}-ray free
  electron lasers in {\em X-ray free electron lasers}, eds.{} S Boutet, P
  Fromme, M Hunter.
\newblock (Springer, Cham), p. 441 (2018).

\end{thebibliography}

\end{document}




\SItext





\subsection*{Start-to-end simulations of the SwissFEL Athos beamline}

Results reported in the main text represent a general, optimized beam dynamics methodology assuming ideal components. Here we also describe start-to-end simulations of the phase-locked ultra fast X-ray (PHLUX) pulse working principle using existing particle tracker modules, and assuming the effective components and specific layout of the SwissFEL Athos soft X-ray beamline \cite{Abela19} at a photon energy of $E\sim1095$~eV. 

The electron bunch is extracted from the linear accelerator at an energy of 3.1~GeV and transported through a switchyard to the undulator beamline. The beam manipulation (tilt) needed for the PHLUX scheme, is applied in the last dispersive region around the position of the switchyard. There, the dispersion is $\sim23$~cm and the electron beam features a residual linear chirp of 1.1\% full-width-at-half-maximum (FWHM) arising from the last compression stage. This yields a transverse beam extension of $\sim2.5$~mm and a correlation between the transverse and longitudinal position that is sufficient to apply higher order tilts with multipole magnets. In the following, we consider octupole magnets as these are currently the highest-order multipole magnets supported by \textit{Elegant} \cite{Borland00}. We note that instead of the `hat'-like profile shown in Fig.~{\color{blue}1}(c), this results in a `step'-like profile where either the head or the tail of the electron bunch contributes to the first and second free-electron laser (FEL) stage---compared to the sextupole and decapole solution this only limits the maximum time delay between the pulses. Beside an emittance spoiler foil (slit width down to 50~$\mu$m optimized for shortest pulse length), which is applied `manually' to the particle tracking, the considered lattice features two octupole magnets of length $L=30$~cm close to the innermost quadrupole pair in the dispersive region. The integrated strengths of the first and second octupole are $k_3L =  45$'000 and $-30$'000 m$^{-2}$, respectively. This corresponds to magnetic fields of up to $B\approx0.25$~T at the octupole poles assuming a bore radius of 1~cm, which is feasible. The lattice requires two octupoles with opposite polarity to reduce the impact on the current profile while applying the third order tilt to the electron bunch. In addition, one of the inner quadrupoles is detuned by $\sim2$\% to add the linear tilt for the finally desired transverse beam shape.

Following the dispersive section with the two octupoles and the emittance spoiler foil, there is an 8~m long `dechirper' to remove the residual energy chirp. We note that this component is part of the baseline design of the Athos beamline since the intermediate outcoupling at 3.1~GeV lacks the wakefields of the additional radio-frequency structures of the Aramis hard X-ray beamline which remove such chirp there. The gap of the dechirper is 2~mm and has a corrugation periodicity and depth of 500 and 250~$\mu$m, respectively. The baseline undulator line starts $\sim90$~m after the switchyard with $\sim40$~m free space in front. A future upgrade of the Athos beamline, relevant for the PHLUX mode, is the implementation of self-seeding, where the first stage will be placed in front of the currently existing undulators. Due to the unique layout of SwissFEL Athos the filtering via self-seeding will utilize an arm of $\sim15$~m, allowing to realize the required high resolving power (see \textit{Discussion on the Self-Seeding Resolving Power}).

\begin{figure}
	\centering
    \captionsetup{width=0.5\linewidth}
	\includegraphics[width=0.5\linewidth]{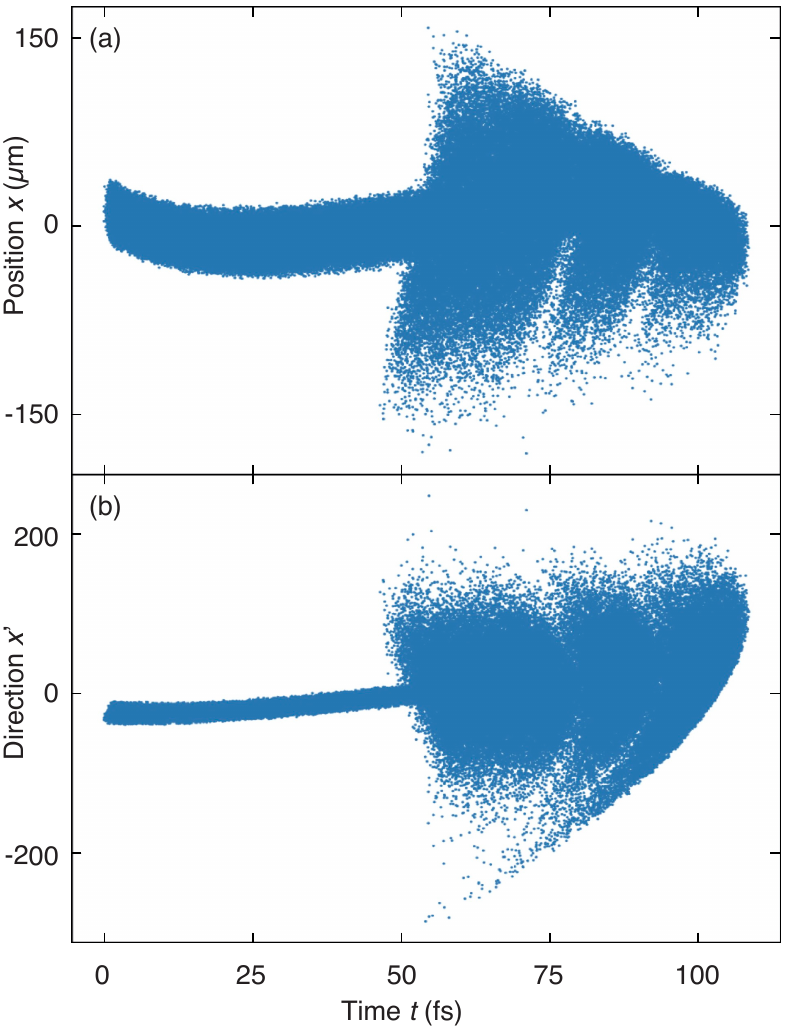}{}
	\caption{\label{phasespace}
		Particle distribution along the electron bunch at the undulator entrance, shown in terms of the horizontal (a) position $x$ and (b) instantaneous direction $x'=dx(z)/dz$. $z$ is the longitudinal coordinate.}
\end{figure}

\begin{figure}
    \centering
    \captionsetup{width=0.5\linewidth}
	\includegraphics[width=0.5\linewidth]{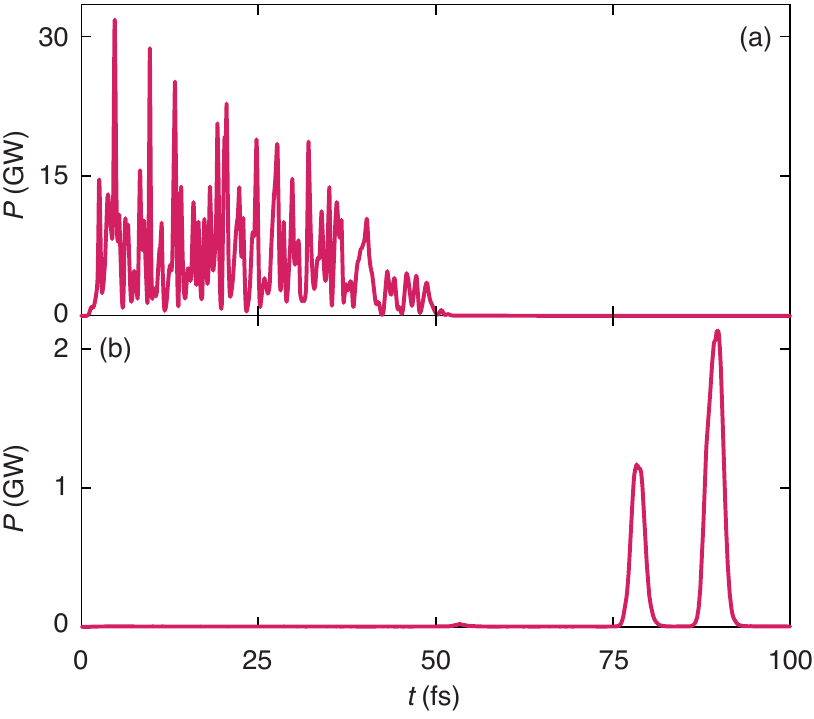}{}
	\caption{\label{profile}
		Radiation power profile at the exit of the (a) first (SASE) and (b) second (seeded) stage. In the latter, the realigned electron bunch and a seed signal of $\sim1$~MW from the monochromator generates a $\sim11$~fs delayed pulse pair with $\sim2.5$~fs FWHM duration.}
\end{figure}

\begin{figure}
    \centering
    \captionsetup{width=0.5\linewidth}
	\includegraphics[width=0.5\linewidth]{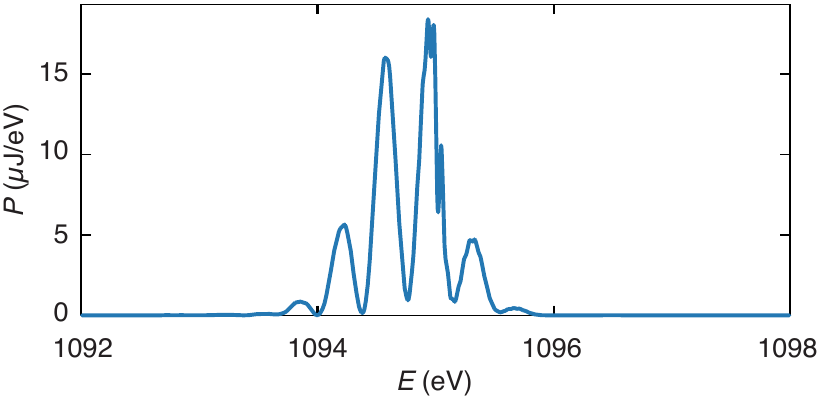}{}
	\caption{\label{spectrum}
		Radiation power spectrum at the exit of the second stage. Here photons are extracted at the point of saturation to increase the tolerance, e.g. with respect to orbit offset and energy detuning.}
\end{figure}

Figure~\ref{phasespace} shows the \textit{Elegant} output of the time-resolved phase space distribution at the undulator entrance in terms of the horizontal position $x$ and instantaneous direction $x'$. Dominant is the blown-up emittance in the first half of the electron bunch, except for two slices from the slits of the emittance spoiler foil. Superimposed is a third-order tilt of both $x$ and $x'$. In the first FEL stage, the beam is aligned to the turning point of the tilt (tail of the bunch). This particle distribution is then imported in \textit{Genesis} \cite{Reiche99} for the two-stage simulation. At the end of the first (SASE) stage, the second half of the electron bunch exhibits the characteristic spiky SASE radiation profile [see Fig.~\ref{profile}(a)]. The applied tilt is sufficient to suppress any lasing in the first half, such that it can be considered a fresh bunch in the second (seeding) stage. In between the two stages, the electron beam is realigned to the head of the bunch and a delay of $\gtrsim1$~ps is applied to wash-out any induced microbunching from the SASE process in the first stage. The second stage is then seeded with the coherent signal of $\sim1$~MW and generates the desired pulse pair [see Fig.~\ref{profile}(b)]. For the \textit{Elegant} simulations slit widths of 50~$\mu$m were assumed to produce short pulses of $\sim2.5$~fs FWHM. Narrower slits would degrade the quality of the electron slice since the width would be smaller than the beam size of the electron bunch at the emittance spoiler foil. Figure~\ref{spectrum} shows the radiation power spectrum exhibiting the expected interference pattern from the pulse pair. Though, some residual from the seed signal is still noticeable, the interference is in good approximation constructive, indicating negligible `phase slippage' due to uneven amplification of the two slices. In the simulation, we assume a fully coherent seed signal over the separation of the two slices. Therefore, the relative phase difference between the two output pulses and, thus, the interference in the spectrum remain stable even if the global phase of the seed signal varies from shot to shot.

These start-to-end simulations demonstrate the feasibility of the PHLUX mode at the example of the layout of the SwissFEL Athos beamline. The next step will be a careful optimization (e.g. balancing of the spikes, scanning of the time separation), as well as improving chromatic effects (beam profile) and embedding decapole magnets in \textit{Elegant} which, however, is beyond the scope of the proof-of-principle calculation reported here.

\subsection*{Discussion of the self-seeding resolving power}

For the self-seeding efficiency, we assume a resolving power of 50'000 such that the seed signal is coherent over the 100~fs length of the electron bunch. Recent work~\cite{Nam21} indeed reports resolving powers of even above 50’000 in the hard X-ray regime where the PHLUX scheme is also applicable. We note that a self-seeding chicane with such a high resolving power is currently not implemented at SwissFEL Athos but foreseen as an upgrade in the $\sim40$~m free space in front of the existing undulators, yielding an arm of $\sim15$~m and allowing to realize the required high resolving power. For further improvement, also schemes using harmonic radiation could be used \cite{Prat18}.

As a comparison, the resolving power of the LCLS soft X-ray self-seeding experiment using a 4~m long break \cite{Ratner15} yields coherence times of $10-20$~fs and, therefore, would not fulfill the requirements for phase-locked double pulses with a separation of more than 50~fs. Considering the incident power on the self-seeding chicane is crucial. Using existing self-seeding and monochromator designs one will be able to operate below the damage threshold. Also, heat dissipation will not be an issue at SwissFEL due the 100 Hz repetition rate. However, we note that to achieve high-resolving powers of order 50’000, taking into account the space constraints in the beamline, one may choose not to drive the first stage fully in saturation. This could influence the stability of the mode due to shot noise from the micro bunch instability, beam jitter etc., but would not compromise the general working principle.

\subsection*{Comparison to other FEL modes}

Table~\ref{comparison} compares the PHLUX scheme to other, more standard operation modes. Namely, conventional self-seeding leads to similar radiation power levels of $\sim$~GW, but pulses are $\sim100 $ times longer. We note that the latter method comes along with a broad ‘pedestal’ background in the spectrum that reduces the coherent fraction of the self-seeding spike, where the integrated area of the pedestal depends on the exact conditions in the accelerator.

The scheme proposed by Thompson and McNeil \cite{Thompson08} delivers trains of few-femtosecond delayed X-ray pulses and builds on the availability of a modulation laser and magnetic chicanes between undulator modules. PHLUX, on the other hand, requires a self-seeding chicane with a very high resolving power and higher-order beam tilts which also has not yet been demonstrated. In return, it allows producing two or more phase-locked pulses with large time delays of up to $\sim100$~fs.

\subsection*{Benchmarking against frequency-domain RIXS}

In recent years, resonant inelastic X-ray scattering (RIXS) has become a powerful experimental technique for spectroscopy of molecular systems \cite{Beye13} and quantum matter \cite{Ament11} via mapping of the energy and momentum transfer relations, where elemental selectivity is achieved by tuning the X-ray energy to particular elemental absorption edges. The inelastic cross-section is related to the dynamic structure factor, which at synchrotrons is measured using sophisticated spectrometers \cite{Strocov10,Huang18,Jarrige18,Brookes18} that, however, feature low collection efficiencies and energy resolutions of tens of meV at best. 

As depicted in Fig.~{\color{blue}4}(b), frequency-domain RIXS employs high-resolution gratings before and after the sample to select the incoming and outgoing photon energy, respectively. Meanwhile, time-domain X-ray interferometry (XRI) only requires at most one dispersive element to analyse the emission spectrum and allows for use of large area detectors which both increase the measurement efficiency and, at the same time, substantially reduce the complexity of instruments. In addition, by changing the sample-detector distance and maximum time delay $\Delta t$ versus time step size, within the Fourier limits XRI allows for independent tuning of momentum and energy resolution, respectively. 

For time-domain XRI measurements, photons are delivered to the sample in a fixed spectral bandwidth $\Delta\omega_{\rm range}\sim1/\tau_{\rm pulse}$. A measurement at $n$ different time delays $\Delta t$ yields a spectral resolution of $\Delta\omega_{\rm res}\sim1/(n\Delta t)$. The signal is then proportional to the number of photons deposited in a resolution bin of width $\Delta\omega_{\rm res}$, corresponding to the number of photons deposited per step $\Delta_t$. 

For frequency-domain measurements, photons are delivered in $n$ steps within a narrow bandwidth $\Delta\omega_{\rm res}$ by scanning a  central frequency $\omega_c$ over a frequency range $\Delta\omega_{\rm range}$. The signal is then proportional to the number of photons deposited on the sample per step $\omega_c$.

Therefore, the parameters relevant for benchmarking are the number of photons per unit time within the bandwidth $\Delta\omega_{\rm res}$ for frequency-domain measurements and the number of photons per unit time in PHLUX pulse pairs for time-domain XRI.
Assuming a resolving power $\omega_0/\Delta\omega_{\rm res}$ of 50'000, about $\sim10^{10}$ photons are delivered by a PHLUX pulse pair, resulting in $\sim10^{12}$ photons per second at the 100 Hz repetition rate of SwissFEL~\cite{Abela19}. At a third generation synchrotron the numbers are of similar magnitude, e.g. the ADRESS beamline at the Swiss Light Source \cite{Strocov10ii} provides about $\sim2\cdot10^{12}$ photons per second at the same resolving power. This shows that XRI at a Cu-based low-repetition rate X-ray FEL roughly breaks even with respect to conventional synchrotron-based RIXS. In turn this also implies that the scheme can be game changing at next-generation superconducting X-ray FELs \cite{Dunn18} where it fully benefits from the kHz to MHz repetition rate. We note that methods other than an emittance spoiler foil may have to be considered at high-repetition rate FELs to reduce the heat load and bremsstrahlung caused by the scattered electrons. For example, slicing of the electron bunch via an optical laser or energy modulation are possible routes.

An additional gain of XRI in terms of efficiency, when used without an analyser (e.g. when either time-dependent speckle or direct amplitude interference terms are being measured), stems from multiplexing in momentum-space which is enabled by the use of large area detectors.

\begin{figure*}
    \centering
    \captionsetup{width=0.786\linewidth}
	\includegraphics[width=0.786\linewidth]{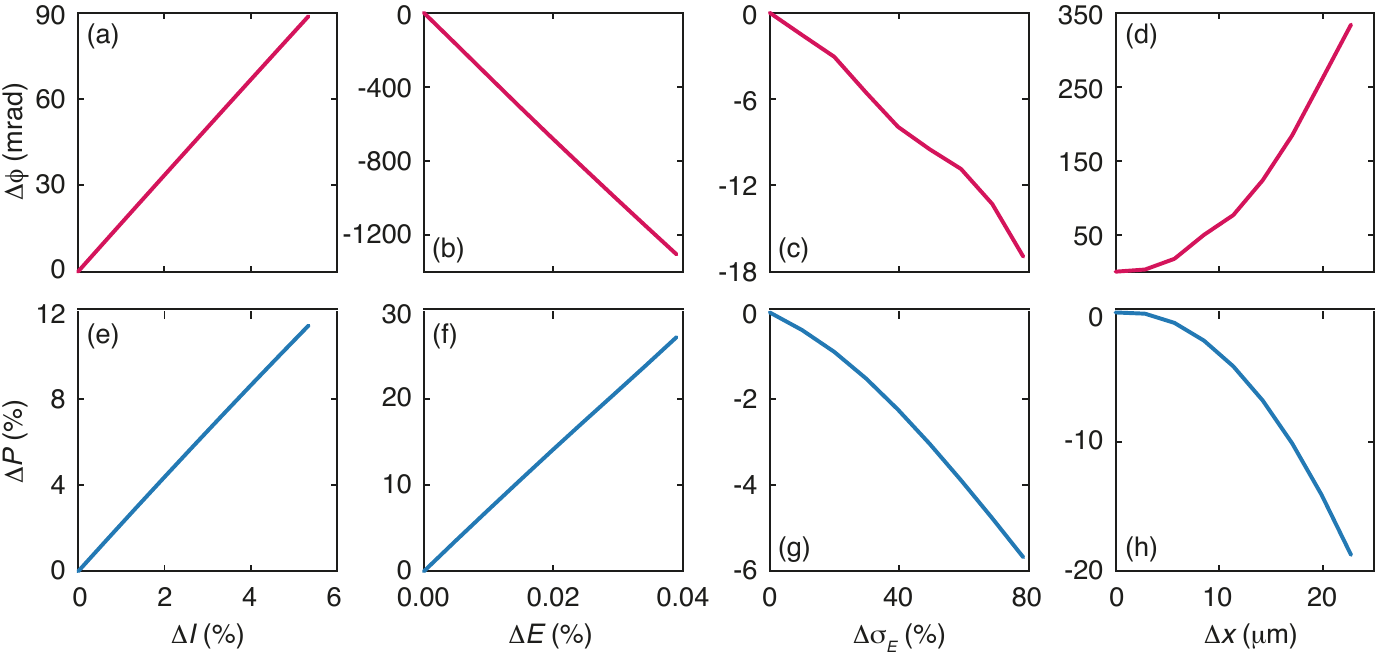}
	\caption{\label{jitter}
		Dependence of the relative phase difference $\Delta\phi$ (a-d) and radiation power $\Delta P$ (e-h) of two phase-locked X-ray pulses upon variation of the current $\Delta I$, mean energy $\Delta E$, energy spread $\Delta\sigma_E$ and axial alignment $\Delta x$ of the electron bunch. For the expected variations at the SwissFEL Athos beamline, the resulting influence on the temporal separation/fluctuations of the two pulses is negligible and rather determined by the mask stability.}
\end{figure*}

\begin{table}
    \captionsetup{width=0.66\linewidth}
    \centering
    \caption{\label{comparison}
    Comparison of the PHLUX, conventional self-seeding and self-amplified spontaneous emission (SASE) operation modes. For the parameters considered here $P_0\sim$~GW and \mbox{$t_0\sim0.5$~fs}.} 
    \begin{tabular}{llll}
        Parameter & PHLUX & Self-seeding & SASE \\ \midrule
        Peak power &        $P_0$ &      $P_0$ &          2$P_0$ \\
        Power fluctuation & small &     small &         100\%, on average $P_0$ \\
        Pulse length (FWHM) &      $t_0$ &    100 $t_0$ &        100 $t_0$, but with $t_0$ spikes \\
        Spectral width &    Fourier limit~ &   Fourier limit~ & Not Fourier-limited, effectively as $t_0$ spikes \\
        Spectral peak intensity~ &   1 & $100^2$ &         100 \\
        \bottomrule
    \end{tabular}
\end{table}

\FloatBarrier






\bibliography{PHLUX}